\documentclass[letterpaper, 10 pt, conference]{ieeeconf}  

\IEEEoverridecommandlockouts                              

\usepackage{hyperref}
\usepackage[english]{babel}
\usepackage{acronym}
\usepackage{amssymb}
\usepackage{amsmath}
\usepackage{graphicx}
\usepackage{url}
\usepackage{hyperref}
\usepackage{epstopdf}
\usepackage{adjustbox}
\usepackage{comment}
\usepackage{makecell}

\acrodef{bci}[BCI]{Brain-Computer Interface}
\acrodef{emg}[EMG]{Electromyography}
\acrodef{eeg}[EEG]{Electroencephalogram}
\acrodef{car}[CAR]{Common Average Reference}
\acrodef{cva}[CVA]{Canonical Variate Analysis}
\acrodef{ic}[IC]{Intentional Control}
\acrodef{inc}[INC]{Intentional Non-Control}
\acrodef{som}[SOM]{Self-Organizing Maps}
\acrodef{smr}[SMR]{Sensory-Motor Rhythms}
\acrodef{mi}[MI]{Motor Imagery}
\acrodef{ros}[ROS]{Robot Operating System}

\title{\LARGE \bf
Entropy-based Motion Intention Identification for Brain-Computer Interface
}

\author{Stefano Tortora$^{1}$, Gloria Beraldo$^{1}$, Luca Tonin$^{2}$ and Emanuele Menegatti$^{1}$
\thanks{$^{1}$Stefano Tortora, Gloria Beraldo and Emanuele Menegatti are with the Intelligent Autonomous Systems Lab (IAS-Lab), Department of Information Engineering (DEI), University of Padova, Italy
        {\tt\small {tortora, beraldog, emg}@dei.unipd.it}}%
\thanks{$^{2}$Luca Tonin is with the Chair in Non-Invasive Brain-Machine Interface, Center for Neuroprosthetics, School of  Engineering, Ecole Polytechnique Fédéral de Lausanne, Lausanne, Switzerland
        {\tt\small luca.tonin@epfl.ch}}%
}

\begin{document}

\maketitle
\thispagestyle{empty}
\pagestyle{empty}

\begin{abstract}
The identification of intentionally delivered commands is a challenge in Brain Computer Interfaces (BCIs) based on Sensory-Motor Rhythms (SMR). It is of fundamental importance that BCI systems controlling a robotic device (i.e., upper limb prosthesis) are capable of detecting if the user is in the so called Intentional Non-Control (INC) state (i.e., holding the prosthesis in a given position). In this work, we propose a novel approach based on the entropy of the Electroencephalogram (EEG) signals to provide a continuous identification of motion intention. Results from ten healthy subjects suggest that the proposed system can be used for reliably predicting motion in real-time at a framerate of 8 Hz with $80\% \pm 5\%$ of accuracy. Moreover, motion intention can be detected more than 1 second before muscular activation with an average accuracy of $76\% \pm 11\%$.
\end{abstract}

\section{INTRODUCTION}
In the last years, thanks to a deeper understanding of the mechanism regulating our brain functions, several \ac{bci} technologies have been developed to allow people with severe disabilities to interact with devices for control or for communication~\cite{dornhege2007toward}. Among those, asynchronous \ac{bci}s based on \ac{smr} have become more sophisticated and reliable both in terms of classification accuracy and speed. Such performances enabled the proliferation of \ac{mi} \ac{bci}s, allowing a more natural control of robotic devices. Recently, researchers started exploiting the principle of "sharing autonomy"~\cite{tonin2010role} in controlling robotic devices, both for mobility~\cite{tonin2011brain} and manipulation~\cite{muelling2015autonomy}. Shared-control approaches allow the user to focus the attention on delivering a reduced number of high-level goals, leaving low level problems to the robotic intelligence (i.e. obstacle avoidance). These solutions have shown to reduce the number of commands needed to navigate a mobile robot through complex environments~\cite{beraldo2018brain}. 

High classification accuracy of \ac{smr}-\ac{bci}s is achieved when the user is intentionally delivering a command (e.g., by discriminating if the user is imagining movements of the hands or the feet). It is also important to design \ac{bci} systems able to recognize when the user does not want to deliver any command, in order not to change the behavior of the robot (i.e., keep an upper limb prosthesis in a given position). This particular situation is known as \ac{inc} state~\cite{satti2009continuous,tonin2017not}. In the past, few solutions have been proposed to face the problem of detecting \ac{inc} state, and they can be principally divided in three groups: i) leaving to the user the burden of actively controlling the \ac{bci} to not deliver any command; ii) exploiting multi-class classification techniques, introducing an extra class that represent the resting state~\cite{zhiwei2007classification,wang2007feature}; iii) adopting control strategies that minimize the delivery of undesired commands~\cite{satti2009continuous,coyle2011eeg}. However all these solutions present some limitations: the first implicates a higher workload for the user to be in control of the robotic device; the second could compromise the classification performance of conventional \ac{smr}-\ac{bci}s due to the difficulty of consistently modelling the unbounded rest class; the third is a promising approach but few people are considering it at present time. \newline
In this paper, we propose a novel approach for motion intention detection based on the entropy of \ac{eeg} signals in order to continuously identify periods of \ac{ic} and Intentional Non-Control (INC). The entropy computed from \ac{eeg} signals have been previously used in several studies to discriminate among different (active) motor imagery tasks~\cite{zhang2008feature,hu2009application,wang2012motor}. However, none of these works faces the challenge of \ac{inc} detection. They all exploit the entropy-based features for conventional \ac{mi} \ac{bci} (i.e., imagining right vs. left hands after a cue) as an alternative to the most common frequency-based features. Herein, we hypothesize the complexity (i.e., entropy) of \ac{eeg} signals might convey orthogonal information about the intention to move. In a previous work~\cite{tonin2017not}, we have shown that an entropy-based \ac{inc} detector coupled to a \ac{smr}-\ac{bci} could reduce the amount of unintentionally delivered commands of about $43\%$ on average, keeping unaltered the performance of the \ac{mi} task. However, in that study, as well as in most of the \ac{mi} studies, the experimental protocol consists of cue-guided discrete trials, where the subject starts at rest and after a certain cue appears has to suddenly start performing the task. In this work we aim to extend the previously proposed system to detect self-paced movements identifying in real-time continuous switching between states of \ac{ic} and \ac{inc} while controlling an external device. The performances of the proposed detector are evaluated in classifying periods of movements execution as indicated by \ac{emg} signals, but also in predicting motion intention before the generation of muscular activity.

This paper is organized as follows. Section II describes the experimental protocol and the BCI system. Section III presents the obtained results, discussed in details in Section IV. Finally Section V provides some conclusions and our future works.
\begin{figure*}
\centering
\includegraphics[width=\textwidth]{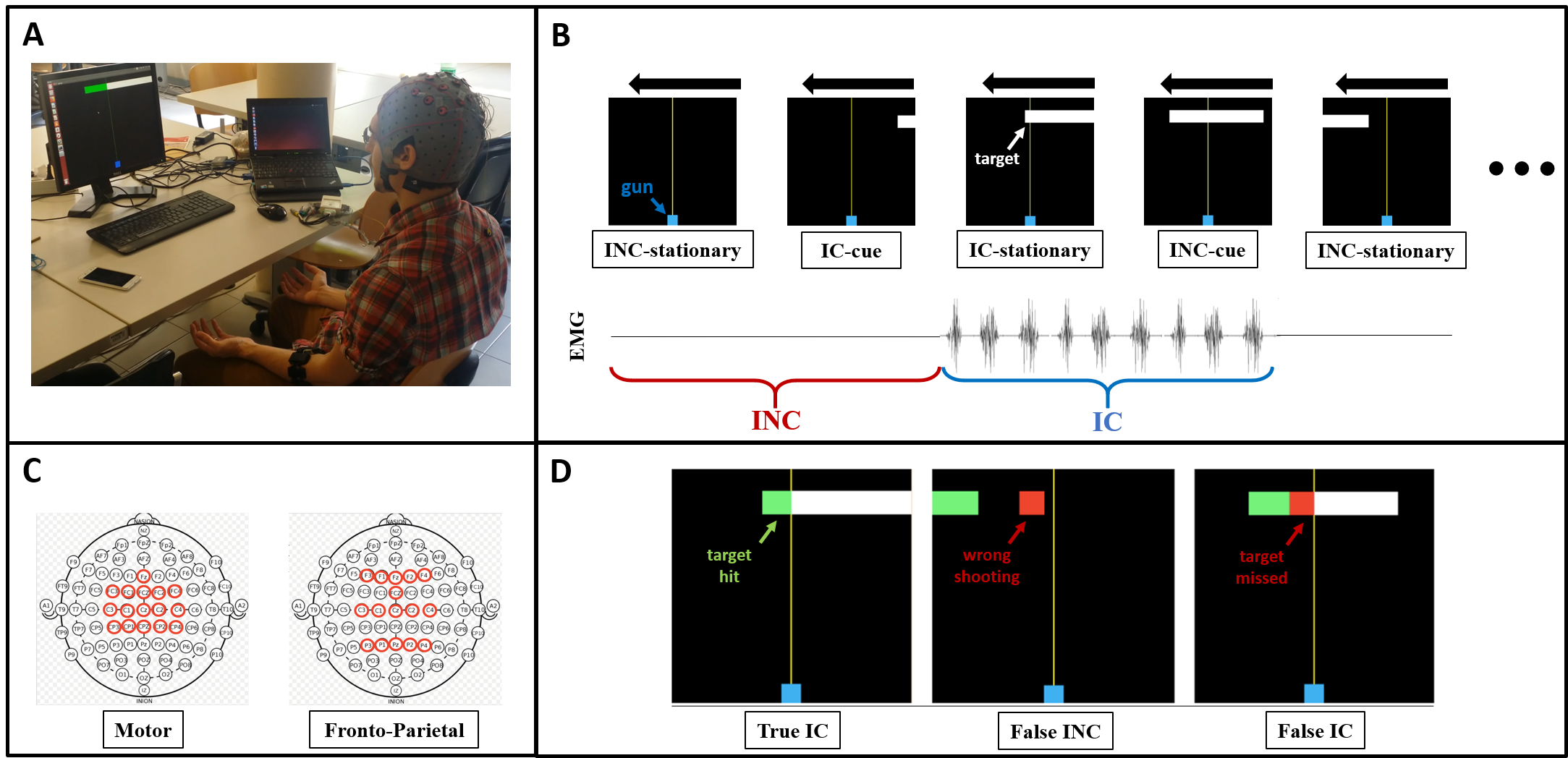}
  \caption{(A) Screenshot of the experimental set up; (B) The protocol consists of a series of 4 consecutive phases (INC-stationary, IC-cue, IC-stationary, INC-cue). INC and IC periods are identified by the absence or the presence of EMG activity, respectively; (C) Channels representation of the two tested configurations; (D) Visual feedbacks provided during the game. The white target turns green when correctly hit (left) or red if missed (right); red rectangles appear when the subject shoots during INC periods (middle).}
\label{fig:protocol}
\end{figure*}
\section{METHODS}
\subsection{Experimental design}
Ten healthy volunteers participated to this study (age from 18 to 58 years old, 4 female). Participants had no known significant health problems at the time of the experiment. All subjects were right-handed and only two subjects (\textit{s02} and \textit{s07}) were familiar with \ac{bci}. The study was carried out in accordance with the principles of the Declaration of Helsinki.

The experimental protocol consisted in a videogame in which subjects were asked to "shoot" to a white rectangle (the target) moving from right to left at a constant speed. Users can "shoot" to the target by means of a stylized gun placed in the middle of the screen. The shooting is controlled through muscular activity (\ac{emg}) of the right and left forearm, by slightly closing the fingers of both hands simultaneously. At the same time, \ac{eeg} signals were acquired and synchronized with the game interface through a custom event sender from \ac{ros} to the \ac{bci} library. A screenshot of the experimental protocol is shown in Figure~\ref{fig:protocol}A. The game is structured in 4 consecutive phases, shown in Figure~\ref{fig:protocol}B and following explained:
\begin{itemize}
    \item \textbf{INC-stationary:} no target is present on the screen or the target has just left the middle of the screen. Subjects are asked to not perform any movement;
    \item \textbf{IC-cue:} the left edge of the target appears on the screen, but it has not reached the middle of the screen yet. Subjects are asked not to perform any movement, but they know they should start moving in few seconds;
    \item \textbf{IC-stationary:} the target reaches the middle of the screen and subjects are asked to shoot to the rectangle, continuously opening and closing both hands;
    \item \textbf{INC-cue:} the right edge of the target appears on the screen, but it has not reached the middle of the screen yet. Subjects are asked to continue shooting to the target, but they know the should stop moving in few seconds.
\end{itemize}
Visual feedbacks are provided to inform subjects if they are behaving correctly, as shown in Figure~\ref{fig:protocol}D: the target turns to green if hit when it is required (IC-stationary, INC-cue) or to red if missed; small red rectangles appear if subjects move while not required (INC-stationary, IC-cue). Each cue condition (IC-cue, INC-cue) lasted $3.0$ seconds, while stationary conditions (IC-stationary, INC-stationary) lasted a random time between $3.5$ and $4.5$ seconds.

Each run of the game consists of the $10$ trials of each condition. Before starting the experiment, each subject tried one run to familiarize with the game. Then, five runs were performed for each subject, resulting in a total of $200$ trials per subject.

\subsection{EMG recording}
The \ac{emg} signals were acquired through two Myo Gesture Control Armbands, by Thalmic Labs. Both Myo sensors have been connected and synchronized to \ac{ros} by means of an open source custom software library\footnote{The library code is publicly available at~\url{https://github.com/ste92uo/ROS_multipleMyo}}. The envelop of \ac{emg} signals has been extracted online by means of rectification and smoothing to control the game interface and provide closed-loop feedbacks of subject's movements. In order to shoot, the \ac{emg} envelops of both forearms should overcome a threshold, manually tuned for each subject in order to have a highly sensitive interface. For classifier training, the \ac{emg} signals have been used to segment the acquired data in time periods when the subject is intentionally controlling the interface generating \ac{emg} activity and time periods when the subject is not contracting the muscles in order to voluntary not control the interface. Thus, data acquired in the cue phases have been merged with data from their corresponding stationary phases, as shown in Figure~\ref{fig:protocol}B--i.e. IC-cue with INC-stationary and INC-cue with IC-stationary. 

\subsection{EEG recording and preprocessing}
\ac{eeg} data were recorded with an active 16-channels acquisition system (g.tec medical engineering, Austria) at $512$ Hz with reference on the right ear and ground on AFz electrode. Two electrodes configurations were considered: i) a "Motor configuration" (Mc) in Figure~\ref{fig:protocol}C (left), that corresponds to the one used in our previous studies~\cite{beraldo2018brain,tonin2017not}, covering the sensory-motor cortex (Fz, FC3, FC1, FCz, FC2, FC4, C3, C1, Cz, C2, C4, CP3, CP1, CPz, CP2, CP4); ii) a "Fronto-Parietal configuration" (FPc) in Figure~\ref{fig:protocol}C (right), with electrodes more spread over the cortex, covering also frontal and parietal regions (F3, F1, Fz, F2, F4, FCz, C3, C1, Cz, C2, C4, P3, P1, Pz, P2, P4). At admission, the participants were randomly assigned to one of these two configurations. Five subjects performed the experiment with one configuration and the remaining five with the other configuration. \ac{eeg} signals from each channel were filtered with a digital notch filter at $50$ Hz by the amplifier, to remove the power line interference, and a \ac{car} spatial filter was applied.

\subsection{Entropy extraction}
The entropy analysis has been performed offline on Matlab, simulating an online \ac{bci} loop with real-time constraints. The recorded data were segmented with a sliding window of $1.5$ seconds and an overlap of $0.125$ seconds, resulting in a framerate of $8$ Hz. Six different frequency bands have been extracted from the filtered \ac{eeg} signals (8-13, 14-22, 22-30, 30-45, 2-45 and 8-30 Hz) with a fourth-order zero-lag Butterworth band pass filter. 
Subsequently, the band-power in each frequency band has been calculated as the envelop of the Hilbert transform of the signal. The Shannon Entropy~\cite{sabeti2009entropy} has been computed for each frequency band. For continuous signals, the Shannon Entropy $H_{sh}$ is defined as
\begin{equation}
    H_{sh}=-\sum_{i}{P_{i}\log{P_{i}}}
\end{equation}
where $i$ ranges over all amplitudes of the signal, $P_i$ indicates the probability of the signal having amplitude $a_i$. For discrete signals, the Shannon Entropy can be computed by means of histogram, linearly dividing the amplitude range of the signal into \textit{k} bins. Finally, the normalized Shannon Entropy is used, computed as
\begin{equation}
    nH_{sh}=\frac{H_{sh}}{\log{k}}
\end{equation}
In this work, $k=32$ has been taken.

The most discriminative features, defined as channel-band pairs, to distinguish between \ac{ic} and \ac{inc} states have been found for each subject by means of \ac{cva}~\cite{galan2007feature}. In this study, \ac{cva} has been used to extract the canonical discriminant spatial patterns (CDSP), representing the features whose directions maximize the difference in mean Shannon Entropy between the classes. The channel-band pairs that contributed the most on the CDSP are identified through a \textit{Discrimimation} index, computed from the pooled correlation matrix between each feature and the CDSP.

A statistical Gaussian classifier was used to compute the posterior class probabilities of each frame; i.e. the probability that at the current frame, the computed entropy over the $1.5$ seconds time window belongs to each class (\ac{ic} or \ac{inc}). Each class is represented by $N_p$ Gaussian prototypes with equal weight $1/N_p$, thus the class-conditional probability density function $p_t(x|y_k)$ for the class $y_k$ at time $t$ can be computed as the superimposition of the Gaussian prototypes. The posterior probability of each class is
\begin{equation}
    p_t(y_k|x)=\frac{p_t(x|y_k)p(y_k)}{\sum_j{p_t(x|y_j)p(y_j)}}
\end{equation}
All classes are assumed to have equal prior probabilities $p(y_k)$. The Gaussian prototypes were initialized by clustering algorithm, namely \ac{som}~\cite{kohonen1990self}. Then, the initial estimation is iteratively improved by means of stochastic gradient descent minimizing the mean square error $E = (1/2)\sum_i{(y_i-t_i)^2}$, where $y_i$ is the predicted class and $t_i$ is the target class. More details can be found in~\cite{millan2004noninvasive}.

In order to smooth the outcomes of the classifier, an exponential integrator has been used to accumulate the evidence over time with the following decision making formula
\begin{equation}
    D(y_t)=\alpha \cdot D(y_{t-1}) + (1-\alpha) \cdot p(y_t|x_t)
\end{equation}
where $D(y_t)$ is the previous level of decision making and $\alpha$ is the integration parameter. Finally, the decision on the predicted class is made by a thresholding procedure with hysteresis. In this sense, if the actual decision level $D(y_t)$ overcomes a threshold $th$, the subject is considered in the state of \ac{ic}; if the decision level goes below $1-th$, the subject is considered in the state of \ac{inc}; otherwise, if the decision level is between $th$ and $1-th$, the previous decision is kept.

\begin{table*}
\caption{Single sample classification accuracy for each subject in classifying between IC and INC states. Comparison between the proposed entropy-based method and the PSD-based method taken from~\cite{millan2004noninvasive}.}
\label{tab:results}
\large
\centering
\begin{adjustbox}{max width=\textwidth}
\begin{tabular}{l | cccccccccc | c}
 & \textbf{s01} & \textbf{s02} & \textbf{s03} & \textbf{s04} & \textbf{s05} & \textbf{s06} & \textbf{s07} & \textbf{s08} & \textbf{s09} & \textbf{s10} & \textbf{Avg}\\
 \hline \hline
\rule{0pt}{3ex}
& \textit{'FPc'} & \textit{'FPc'} & \textit{'FPc'} & \textit{'Mc'} & \textit{'Mc'} & \textit{'Mc'} & \textit{'Mc'} & \textit{'FPc'} & \textit{'FPc'} & \textit{'Mc'} & \\
\hline
\rule{0pt}{3ex}
\textbf{Entropy} & \textbf{0.83} & 0.74 & \textbf{0.83} & 0.83 & \textbf{0.73} & \textbf{0.76} & 0.75 & \textbf{0.85} & \textbf{0.85} & \textbf{0.82} & \textbf{0.80 $\pm$ 0.05} \\
\hline
\rule{0pt}{3ex}
\textbf{PSD} & 0.74 & \textbf{0.78} & 0.80 & \textbf{0.85} & 0.60 & 0.69 & \textbf{0.78} & 0.84 & 0.82 & 0.76 & 0.76 $\pm$ 0.07 \\
\hline
\end{tabular}
\end{adjustbox}
\end{table*}
\section{RESULTS}
\subsection{Motion execution detection (IC vs. INC)}
\begin{figure}
\centering
\includegraphics[width=\columnwidth]{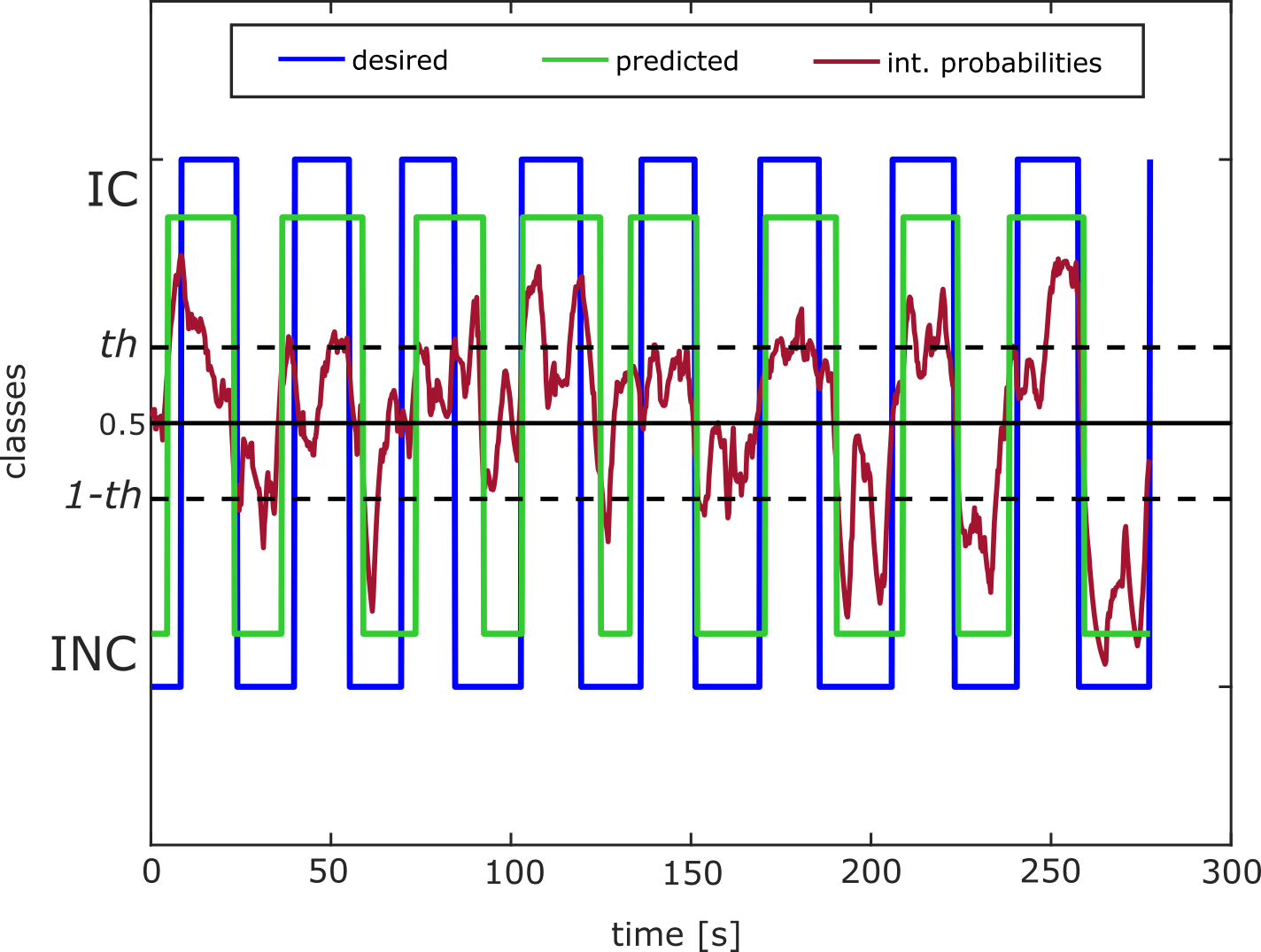}
  \caption{Second testing session of subject \textit{s01}. The classifier outputs is represented by the integrated probabilities (red line). Integrated probabilities higher than 0.5 correspond to IC state and lower to INC state. When the integrated probabilities cross the thresholds \textit{th} or \textit{1-th} (black dotted lines) the IC or the INC classes are predicted respectively (green line), in order to follow the desired classification outputs (blue line).}
\label{fig:ipp}
\end{figure}

The first three runs of each subject has been used as training dataset for features selection and classifier training. Performances have been evaluated on the last two sessions. Table~\ref{tab:results} shows the performance for each subject of the proposed entropy-based \ac{bci} in detecting periods of movement execution (\ac{ic}) from periods of resting (\ac{inc}). For comparison, the same dataset has been also classified by a conventional \ac{smr}-\ac{bci} based on the PSD of the \ac{eeg} signals~\cite{millan2004noninvasive}. From the results, our system achieved comparable performance and better classification accuracy for 7 out of 10 subjects, with an average improve of $3.3\% \pm 5.5\%$, even if not statistically significant (t-test, p$>$0.05). The group of subjects belonging to the \textit{'FPc'} configuration shows on average slightly better performances compared to subjects wearing the \textit{'Mc'} configuration, with classification accuracy of $82.0\% \pm 4.6\%$ and $77.0\% \pm 4.4\%$ respectively for the entropy-based \ac{bci}, and $79.0\% \pm 3.8\%$ and $73.0\% \pm 9.5\%$ respectively for the PSD-based \ac{bci}, but with no statistically significant difference (t-test, p$>$0.05). It is worth mentioning, that the results refer to the single sample accuracy of the testing sessions, thus providing a classification output at a rate of 8 Hz, as shown in Figure~\ref{fig:ipp}. The red line represents the posterior probabilities of the \ac{ic} state after the exponential integration. When the integrated probabilities crosses the thresholds (black dotted lines), the corresponding prediction is generated (green line). The integration parameter $\alpha$ and the decision threshold $th$ were identified for each subject by means of discrete grid search and the best pair of parameters ($\hat{\alpha}$, $\hat{th}$) are selected as the one maximizing the prediction accuracy in a 3-fold cross-validation over the training dataset. $\hat{\alpha}$ equal to 0.9 and $\hat{th}$ equal to 0.65 has been selected for almost all the subjects (8 out of 10 subjects). The results show the feasibility of using the proposed entropy-based \ac{bci} for real-time detection of intentional movements execution locked to the \ac{emg} activity (blue line) with maximum accuracy of $85\%$ for \textit{s08} and \textit{s09} and minimum accuracy of $73\%$ for \textit{s05}.
\subsection{Topographic distributions}
\begin{figure*}
\centering
\includegraphics[width=\textwidth]{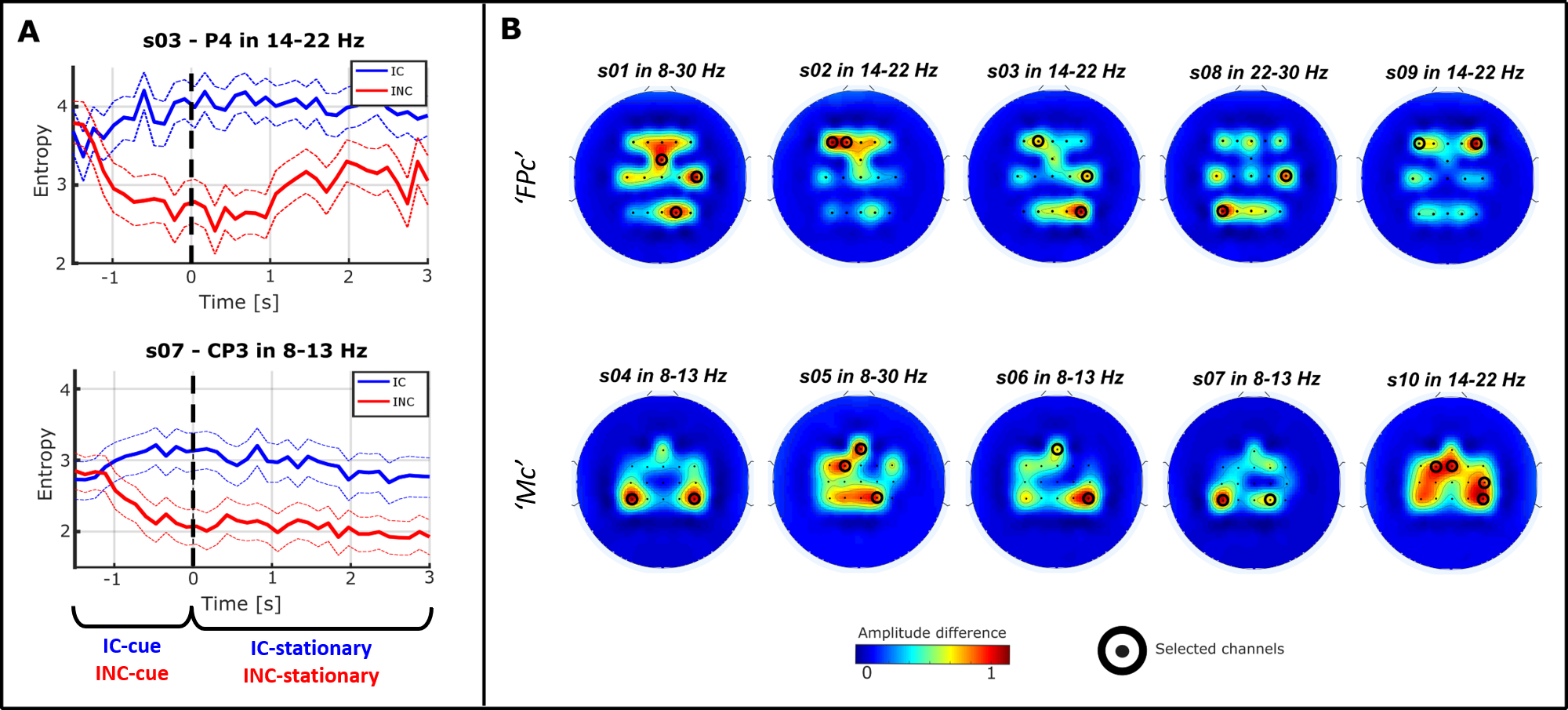}
  \caption{(A) Entropy dynamics average across trials for sample subjects with the \textit{'FPc'} (top) and \textit{'Mc'} (bottom) channel configurations. The entropy shows to increase when the subject is in IC states (blue line), while to decrease during INC states (red line); (B) Topographic maps of the average difference of the entropy amplitude between the IC and INC states for subjects with the \textit{'FPc'} (top) and \textit{'Mc'} (bottom) channel configurations. The selected frequency bands correspond to the most discriminant bands for each subject. The channels selected by CVA for each subject in the represented frequency band are highlighted (circled dots), showing a higher contribution of frontal and parietal areas in classifying among IC and INC states.}
\label{fig:topoplot}
\end{figure*}
\begin{figure}
\centering
\includegraphics[width=\columnwidth]{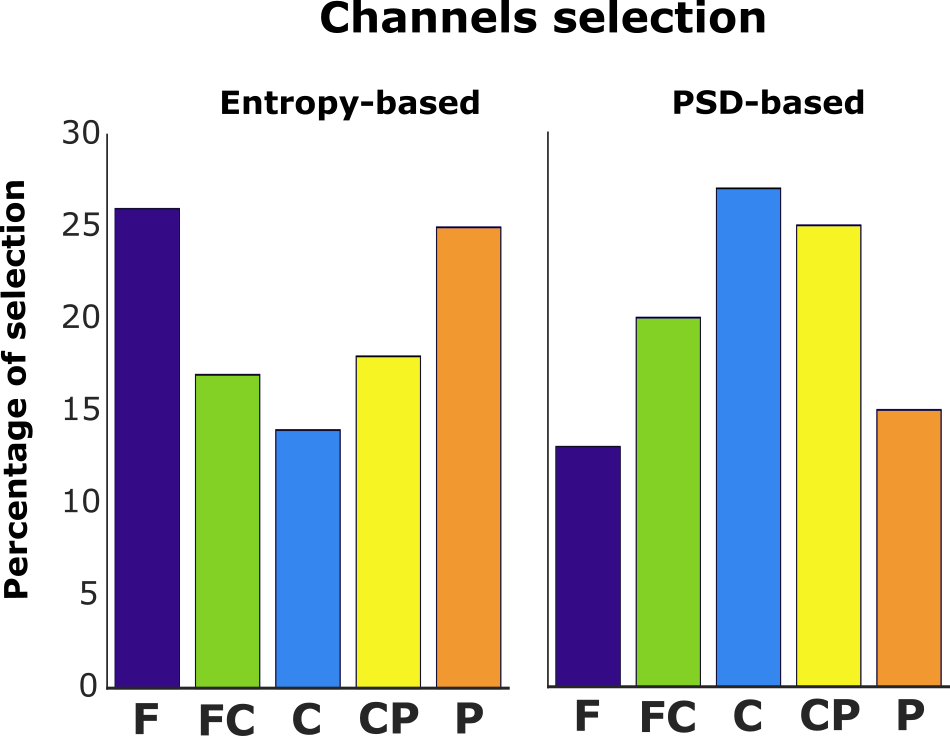}
  \caption{Channels selection distribution over the cortex for all subjects. Selection based on entropy features (left) shows a higher importance of channels from the frontal (F) and parietal (P) areas, while selection based on PSD features (right) from the central (C) and centro-parietal (CP) areas.}
\label{fig:chan_sel}
\end{figure}
As expected from the results in~\cite{tonin2017not}, the intentional execution of a movement (\ac{ic} state) generates an increase of the complexity in the brain activity, as measured by the entropy of the \ac{eeg} signals, compared to resting periods (\ac{inc} state), as shown in Figure~\ref{fig:topoplot}A. Figure~\ref{fig:topoplot}B shows the topographic maps for each subject in the frequency band that has been selected the most by \ac{cva}. Red areas represent locations where the difference between the entropy in the \ac{ic} and \ac{inc} states is higher, and blue areas where no significant difference is evident. Black circled dots represent the channels selected by \ac{cva} in the shown frequency band. Results show that intentional movements execution is discriminated mostly by frontal and parietal electrodes for both configurations, \textit{'FPc'} (top) and \textit{'Mc'} (bottom), except for C4 that has been selected in four out of ten subjects. This tendency has been confirmed by analyzing the percentage of times the electrodes in each line have been selected by \ac{cva} for all the subjects (Figure~\ref{fig:chan_sel}). It can be seen that for the entropy-based \ac{bci}, \ac{cva} selected more than $50\%$ of the features in channels from lines F and P, about $35\%$ in channels from lines FC and CP, and only $14\%$ from the central line (C), even if it was present in both configurations. On the other hand, by applying \ac{cva} on PSD features, more than $50\%$ of the features are selected from channels in the central (C) and centro-parietal (CP) lines, while less than $25\%$ in the frontal (F) and parietal (P) electrodes.

\subsection{Motion preparation detection (INC vs. IC-cue)}
\begin{figure}
\centering
\includegraphics[width=\columnwidth]{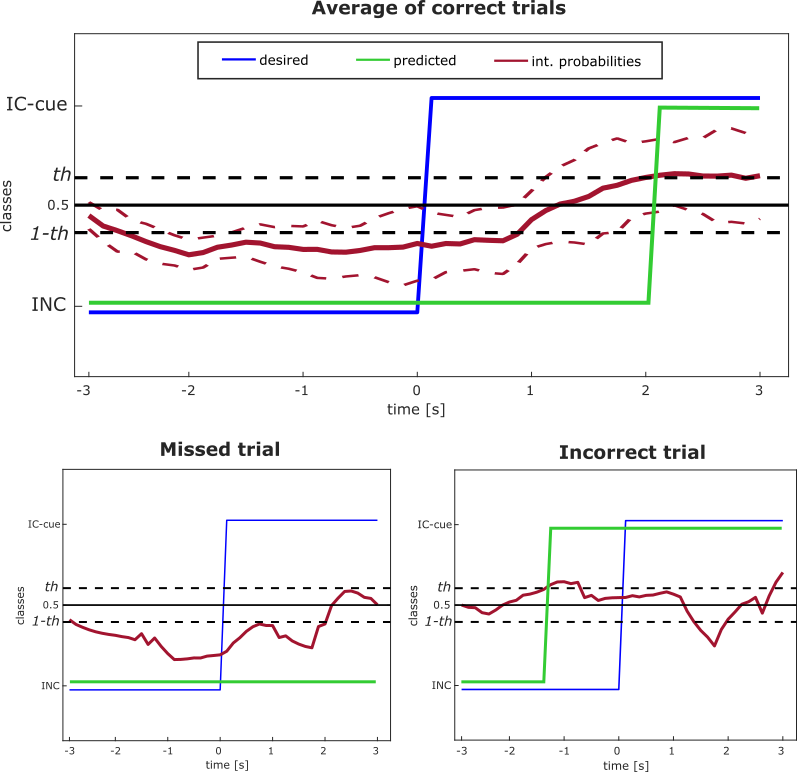}
  \caption{Single trial classification performance in detecting motion intention during IC-cue phase. The figure shows the classifier outputs (red), prediction (green) and the desired classification (blue) for the average of correctly classified trials (top), for one missed trial (bottom left) and one incorrect trial (bottom right) of subject \textit{s04}}
\label{fig:tr_class}
\end{figure}

As shown in Figure~\ref{fig:topoplot}A, the difference between the entropy in the \ac{ic} and \ac{inc} states is evident already during the appearance of their cue phases (IC-cue and INC-cue respectively), and then reaching a plateau from the beginning of their stationary phases. Indeed, from the analysis of the spectrograms of channels related to movements of both hands (C3, C1, C2 and C4) averaged across all trials of every subject, it can be seen an Event-Related Desynchronization (ERD) after the beginning of the IC-cue phase, particularly evident in $\beta$ band (14-22 Hz). Paired t-tests have confirmed a statistically significant difference (p$<$0.0001) between the spectrogram 3 seconds before the IC-cue (\ac{inc} phase) and in the 3 seconds of the IC-cue phase, for all the analyzed electrodes in both $\mu$ and $\beta$ bands. These evidences suggest that motion intention of self-paced movements can be detected from the brain activity before muscular activation. To this aim, the proposed entropy-based \ac{bci} has been trained to classify between \ac{inc} state and IC-cue state. The results are reported for each subject in Table~\ref{tab:results2} for the single sample classification and the single trial classification. The single sample classification accuracy show that the entropy of the \ac{eeg} signals can be used to identify in real-time with a framerate of 8 Hz motion intention before muscles contraction with performance slightly lower respect to the detection of movement execution. The system provides an overall classification accuracy of $74\% \pm 5\%$ with maximum performance of $82\%$ for \textit{s04}. The single trial classification performance are measured as the accuracy and the delay in predicting motion intention after the appearance of the IC-cue. 
Thus, as shown in Figure~\ref{fig:tr_class}, if the system identifies motion intention in the 3 seconds of the IC-cue phase, the trial is considered correctly predicted; if motion intention is predicted in the 3 seconds before the IC-cue phase (INC phase), the trial is considered incorrect; otherwise, if no motion intention is predicted before the end of the IC-cue phase, the trial is considered as missed. The proposed \ac{bci} is able to predict the intention of performing a motion with a single trial accuracy of $76\% \pm 11\%$ in $1.7 \pm 0.7$ seconds on average after the beginning of the IC-cue phase. On average, $15\% \pm 6\%$ of the trials are missed, while less than $9\%$ are classified before the cue, showing the robustness of the system to missclassifications.
\begin{figure*}
\centering
\includegraphics[width=\textwidth]{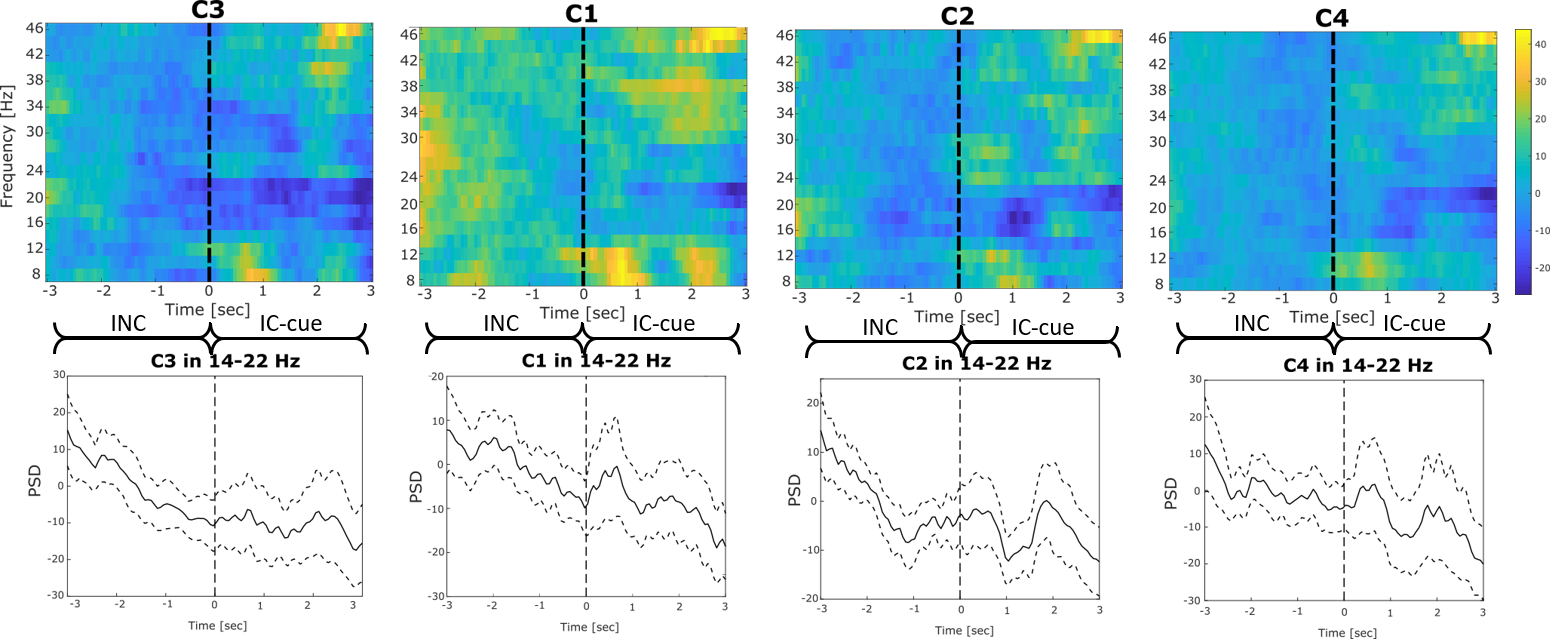}
  \caption{Spectrograms of the EEG signals in central electrodes (C3, C1, C2, C4) baseline corrected and average across all trials of every subject. An Event-Related Desynchronization (ERD) can be seen over all electrodes after 1.5 seconds on average from the beginning of the IC-cue phase, particularly evident in $\beta$ band (14-22 Hz).}
\label{fig:spect}
\end{figure*}
\begin{table}
\caption{Classification performance for each subject in classifying between INC and IC-cue states.}
\label{tab:results2}
\centering
\begin{adjustbox}{max width=\columnwidth}
\begin{tabular}{l | ccccc}
 & \textbf{config} & \makecell{\textbf{Single Sample} \\ \textbf{Accuracy [\%]}} & \makecell{\textbf{Single Trial} \\ \textbf{Accuracy [\%]}} & \makecell{\textbf{Single Trial} \\ \textbf{Missed [\%]}} & \textbf{Delay [sec]} \\
 \hline \hline
\rule{0pt}{3ex}
s01 & \textit{'FPc'} & 0.78 & 0.90 & 0.10 & 1.5 $\pm$ 0.7 \\
\hline
\rule{0pt}{3ex}
s02 & \textit{'FPc'} & 0.66 & 0.75 & 0.20 & 1.6 $\pm$ 0.8 \\
\hline
\rule{0pt}{3ex}
s03 & \textit{'FPc'} & 0.70 & 0.75 & 0.10 & 1.5 $\pm$ 1.2 \\
\hline
\rule{0pt}{3ex}
s04 & \textit{'Mc'} & 0.82 & 0.95 & 0.05 & 1.5 $\pm$ 0.9 \\
\hline
\rule{0pt}{3ex}
s05 & \textit{'Mc'} & 0.75 & 0.70 & 0.25 & 1.9 $\pm$ 0.4 \\
\hline
\rule{0pt}{3ex}
s06 & \textit{'Mc'} & 0.73 & 0.80 & 0.20 & 1.2 $\pm$ 1.0 \\
\hline
\rule{0pt}{3ex}
s07 & \textit{'Mc'} & 0.78 & 0.85 & 0.10 & 1.5 $\pm$ 0.5 \\
\hline
\rule{0pt}{3ex}
s08 & \textit{'FPc'} & 0.77 & 0.65 & 0.20 & 2.6 $\pm$ 0.6 \\
\hline
\rule{0pt}{3ex}
s09 & \textit{'FPc'} & 0.71 & 0.65 & 0.15 & 1.8 $\pm$ 0.3 \\
\hline
\rule{0pt}{3ex}
s10 & \textit{'Mc'} & 0.69 & 0.60 & 0.20 & 1.7 $\pm$ 0.7 \\
\hline
\hline
\rule{0pt}{3ex}
Avg &  & 0.74 $\pm$ 0.05 & 0.76 $\pm$ 0.11 & 0.15 $\pm$ 0.06 & 1.7 $\pm$ 0.7 \\
\hline
\end{tabular}
\end{adjustbox}
\end{table}

\section{DISCUSSION}
As mentioned in the introduction, providing a continuous and robust identification of motion intention is a challenge in \ac{bci} systems. Ideally, the \ac{inc} detector should be able to accurately identify when the user wants to intentionally control the \ac{bci} independently from the specific motor tasks to be performed. Thus, the system should recognize the intention of performing a general motion. The results in this paper strengthen the hypothesis presented in~\cite{tonin2017not} of the existence of a correlation between the complexity of the \ac{eeg} signals, measured by means of Shannon Entropy, and the fact that the user is resting (INC state) or performing a movement-related task (IC state). The Shannon Entropy has been previously used to measure the user's level of attention during \ac{smr} tasks~\cite{saeedi2012real}. In particular, high entropy values are related to higher attention and fast delivery of \ac{bci} commands, while low entropy values to slow delivery of \ac{bci} commands due to reduced attention. From neuroscientific literature~\cite{pfurtscheller2006mu}, we would expect movement-related features in $\mu$ and $\beta$ bands to be more evident right above the motor cortex. This fact has been confirmed also in our study by looking to the channels selected from the PSD features, shown in Figure~\ref{fig:chan_sel} (right). Interestingly, the difference in brain complexity between \ac{inc} and \ac{ic} states is more evident in electrodes from frontal and parietal areas in both configurations as shown in Figure~\ref{fig:chan_sel} (left). On the other hand, only few information is carried out by central electrodes, suggesting that the entropy provides different additional information to motion detection respect to the spectrogram analysis, that could improve the identification of self-paced motion intention, as shown in Table~\ref{tab:results}. Other studies~\cite{brass2007or} have confirmed the role of frontal and parietal areas in the generation of "intentionality" in executing movements. They have shown that the network related to motion intention generation starts from the pre-supplementary motor area (pre-SMA) in the frontal cortex, passes through the precuneous in the parietal cortex up to the pre-motor and motor cortex for movement execution~\cite{haggard2008human}. In this sense, our results suggest that the entropy of the \ac{eeg} signals in frontal and parietal regions could be used as signature of motion intention, independently from the specific motion task. \newline
The first objective of this study was to predict motion execution locked to \ac{emg} activity. According to the results in Figure~\ref{fig:ipp}, the integrated probabilities computed from the entropy-based features robustly follow the dynamic of real movements, continuously identify when the user starts and ends the execution of a movement with time delays in the order of few seconds ($2.5 \pm 3.4$ sec). Thus, the entropy-based \ac{bci} could be used as an alternative to or in conjunction with \ac{emg}-triggered neuroprostheses for people with severe motor disabilities and muscles weakness~\cite{carlson2013hybrid}. \newline
Previous works highlighted that $\beta$ ERD could be associated with motor planning and other functions related to an attentive state of the brain during motor imagery tasks~\cite{molteni2007analysis}. The statistical tests on the spectrogram of the signal before and after the appearance of the IC-cue prove the existence of a self-paced motion preparation prior to movement onset. Results in Table~\ref{tab:results2} shows acceptable classification accuracy of movement preparation respect to resting phases (INC vs. IC-cue), with comparable accuracy to the detection of movement execution (IC vs. INC). It is worth highlighting that, differently from the classification of movement executions that were guided by the dynamics of the game interface (i.e., the subject should start and stop shooting immediately after the left and right edges of the target cross the middle of the screen, respectively), movement preparation was self-paced since the subject could start planning the movement at anytime after the appearance of the IC-cue, making its identification more challenging. These results suggest the feasibility of exploiting the proposed entropy-based \ac{bci} for self-paced motor imagery \ac{bci} applications, since it is able to predict motion intention also when no muscle activity is generated.

\section{CONCLUSIONS}
This study aims to provide a novel \ac{bci} system for continuous prediction of Intentional Control (IC) and Intentional Non-Control (INC) states. The results reveal the possibility to improve the prediction of self-paced motion intention from the complexity of brain activity, measured as the entropy of \ac{eeg} signals, principally in $\alpha$ and $\beta$ bands. The proposed entropy-based \ac{bci} provides comparable performance in detecting motion execution (IC vs. INC) and motion preparation (INC vs. IC-cue), making it feasible to be used for the detection of active motor imagery tasks. Indeed, single trial classification performance shows that motion intention could be reliably predicted more than 1 second before the generation of muscular activity. Moreover, topographic maps in this study and evidences from literature suggest that the analysis of the connectivity between frontal and parietal areas could deep the understanding of the generation of motion intention, independently from the specific motor task to be executed.
In future works, we are planning to evaluate the performance of the proposed method as INC detector in a two-class \ac{smr}-\ac{bci}, in order to reduce the amount of unintentionally delivered commands and improve the control of robotic devices with shared autonomy.

\addtolength{\textheight}{-12cm}   

\bibliographystyle{IEEEtran}
\bibliography{my_biblio}

\end{document}